\documentclass[12pt]{article}
\usepackage{amssymb,amsmath,amsthm}
\usepackage{array,longtable}
\usepackage[dvips]{graphicx}
\usepackage[dvips]{color}
\usepackage[mathscr]{eucal}

\newcommand{\finkfile}{}

\newcolumntype{V}{>{$}m{4cm}<{$}}
\newcolumntype{C}{>{$}c<{$}}
\newcolumntype{L}{>{$}l<{$}}
\newcolumntype{R}{>{$}r<{$}}

\newcommand{\Zh}{\mathbb Z}

\newcommand{\ra}{\rightarrow}

\newcommand{\Ac}{\mathcal{A}}
\newcommand{\Bc}{\mathcal{B}}
\newcommand{\Ic}{\mathcal{I}}

\newcommand{\Qc}{\mathcal{Q}}

\newcommand{\Kc}{\mathcal{K}}

\newcommand{\Uc}{\mathcal{U}}
\newcommand{\Oc}{\mathcal{O}}
\newcommand{\pd}{\partial}

\newcommand{\eps}{\varepsilon}
\newcommand{\Tr}{\mathop{\mathrm{Tr}}\nolimits}
\newcommand{\la}{\langle\!\langle}
\renewcommand{\ra}{\rangle\!\rangle}

\allowdisplaybreaks[3]

\renewcommand{\theequation}{\thesection.\arabic{equation}}

\begin{document}
\title{
\textsc{Construction}\\
\textsc{of the Vacuum String Field Theory}\\
\textsc{on a non-BPS Brane}\\}
\author{
\textsf{I.Ya.~Aref'eva}\\
\emph{Steklov Mathematical Institute,}\\
\emph{Gubkin st. 8, GSP-1, Moscow, 117966, Russia}\\
\texttt{arefeva@mi.ras.ru}\\
\\
\textsf{D.M.~Belov\footnote{On leave from Steklov Mathematical Institute.}}\\
\emph{Department of Physics, Rutgers University,}\\
\emph{Piscataway, NJ 08855, USA}\\
\texttt{belov@physics.rutgers.edu}\\
\\
\textsf{and}\\
\textsf{A.A.~Giryavets}\\
\emph{Faculty of Physics, Moscow State
University,}\\
\emph{Moscow, 119899, Russia}
\\
\texttt{alexgir@mail.ru}
}

\date {~}
\maketitle
\thispagestyle{empty}

\begin{abstract}
In the framework of the Sen conjectures a construction of vacuum superstring
field theory on a non-BPS brane is discussed.
A distinguished feature of this theory is a presence of a ghost kinetic operator
 mixing  GSO$\pm$ sectors.
We proposed a candidate for such kinetic operator with zero cohomology.
\end{abstract}
\newpage
\pagenumbering{arabic}

\tableofcontents


\section{Introduction}\label{sec:intro}
\setcounter{equation}{0}

During the last year the bosonic vacuum string field theory (VSFT)
 proposed to describe physics around  the bosonic tachyon vacuum
  \cite{F2} has been  investigated in many papers
\cite{zwiebach}-\cite{0201159}.
VSFT action has the same form as
the original Witten SFT action \cite{Witten},
but with a new differential operator $\mathcal{Q}$ (for a review of
SFT see \cite{0102085,0109182,ABGKM}). The
absence of physical open string excitations around the tachyon
vacuum \cite{sen-con}-\cite{0105024}
supports a suggestion \cite{F2} that after some field
redefinition $\mathcal{Q}$ can be written as a pure ghost
operator.  Under this assumption
solutions to VSFT equation of motion
admit a  factorized form with the projector-like matter part.
Solutions of projector equation have been discussed in many details in
the references \cite{zwiebach}-\cite{rsz-4}.
This equation is similar to the non-commutative soliton
equations  in the large non-commutativity limit \cite{GMS}.

A generalization of VSFT
to superstrings has been discussed in \cite{F2} and more recently in \cite{AGM}
and \cite{Marino} in the context of cubic SSFT \cite{AMZ1,PTY} and
non-polynomial SSFT \cite{0002211},
respectively.
Open fermionic string in the NSR formalizm has a
tachyon in the GSO$-$ sector that leads to a classical instability of
the perturbative vacuum in the theory without supersymmetry. It
has been proposed  \cite{sen-con} to interpret the tachyon
condensation in the GSO$-$ sector of the NS string as a decay of unstable
non-BPS D9-brane.

The cubic action unifying NS GSO$\pm$ sectors
was constructed \cite{ABKM}
as a generalization of the cubic action for GSO$+$ sector \cite{AMZ1,PTY}.
As in the bosonic case
the vacuum superstring field theory  (VSSFT) is obtained by a shift
of   string field $\hat{A}=(A_+,A_-)$, which describes  both
GSO$+$ and GSO$-$ sectors, by the tachyon vacuum
$\hat{A}_0=(A_{0,+},A_{0,-})$. This shift leads to the
new (shifted) BRST charge, that inevitably has matrix form.
Assuming the statements of Sen conjectures we can also
think that after a proper field redefinition the shifted
BRST charge acts non-trivially only in (super)ghost sector.
In this case VSSFT equations of motion factorizes into
a matter part and a ghost part.
But the factorization in the ghost part is a bit different as compare to one
in the bosonic VSFT.  In the matter part we will have standard
equations for the projectors.
Fermionic projectors, such as the sliver,
have been constructed  recently in \cite{AGM,Marino}.

In the present paper we discuss a possible form of
the vacuum ghost kinetic operator  $\hat{\Qc}$ in cubic VSSFT.
Within  the level truncation scheme it has been  shown \cite{ABKM}
that the tachyon potential in the cubic open  SFT
 has a
non-trivial minimum with $A_{0,+}\neq 0$ and  $A_{0,-}\neq 0$.\footnote{Note, that the pure GSO+ slightly
modified theory  has a non-trivial saddle-point $A_{0,+}\neq 0$
at list at the first truncated levels \cite{AMZ-vac}. However in this case there is
no reason
to assume that the corresponding ${\cal Q}$ is pure ghost.}
Therefore, there is a reason to assume that the vacuum ghost kinetic operator $\hat{\Qc}$:
\begin{equation}
\hat{\Qc}=\begin{pmatrix}
  \Qc_{\textsf{odd}} & \Qc_{\textsf{even}} \\
  -\Qc_{\textsf{even}} & -\Qc_{\textsf{odd}}
  \end{pmatrix},
  \label{QQcal}
  \end{equation}
mixes GSO$\pm$ sectors, i.e.
with $\Qc_{\textsf{even}}\neq 0 $.
If $\Qc_{\textsf{even}}$ would be zero,
we can take $\Qc_{\textsf{odd}}$ to be the ghost kinetic
operator used in the bosonic VSFT \cite{rsz-6}.
   But since it is not the case
   we have to search for another expression for $\Qc_{\textsf{odd}}$.

To be derivatives of the star algebra operators
$\Qc_{\textsf{odd}}$ and $\Qc_{\textsf{even}}$ should satisfy
some identities, in particular, $\langle{\mathcal{I}}|\Qc_{\textsf{odd}}=0$
and $\langle{\mathcal{I}}|\Qc_{\textsf{even}}=0$,
here $\mathcal{I}$ is the identity of the star algebra \cite{GrJe3}.
Both  $\langle {\mathcal{I}}|\Qc_{\textsf{odd}}$  and
$\langle{\mathcal{I}}|\Qc_{\textsf{even}}$ are  singular \cite{GrJe3}, therefore
these expressions  require
 some regularization and
we will discuss it in Section 3.2.

\begin{figure}[!h]
\centering
\includegraphics[width=350pt]{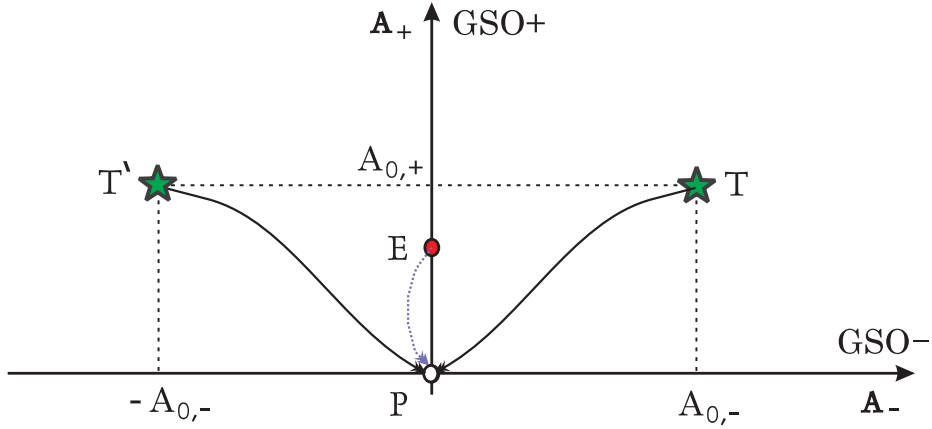}
\caption{Scheme of vacua in SSFT. The point P denotes the perturbative
vacuum, the points T and T$'$ denote the true vacua,
the E denotes a solution found in \cite{AMZ-vac}.
The arrow connecting $T$ and $P$ denotes a solution of VSSFT representing
the original non-BPS brane.}
\label{fig:1}
\end{figure}

Note that to come back from VSSFT to the perturbative vacuum
one has to study
solutions of the  VSSFT equations
with non-zero components in both GSO$+$ and GSO$-$ sectors
(see Figure~\ref{fig:1}).

\bigskip
The paper is organized as follows. In Section~\ref{sec:action} we describe
cubic vacuum superstring field theory on the non-BPS $\mathrm{D}$-brane.
Construction of the kinetic operators is presented in Section~\ref{sec:newBRST}.
Appendix~\ref{app:A} contains some technical details about matrix representation
of the cubic action. Appendix~\ref{app:B} contains the construction
of non-polynomial Vacuum SSFT on a non-BPS brane.


\section{Cubic Vacuum String Field Theory on a non-BPS $\mathrm{D}$-brane }\label{sec:action}
\setcounter{equation}{0}
\subsection{Review of Cubic String Field Theory on a non-BPS $\mathrm{D}$-brane}
\setcounter{equation}{0}

To describe the open string states living on a
single non-BPS $\mathrm{D}$-brane one has to consider
GSO$\pm$ states \cite{sen-con}.
GSO$-$ states are Grassmann even, while
GSO$+$ states are Grassmann odd (see Table \ref{tab:1}).
The unique (up to rescaling of the fields)
gauge invariant cubic action unifying GSO$+$
and GSO$-$ sectors is \cite{ABKM}
\begin{equation}
\begin{split}
S[A_+,A_-]&=\frac{1}{g^2_o}\left[
\frac{1}{2}\langle\!\langle Y_{-2}|A_+,Q_BA_+
\rangle\!\rangle+\frac{1}{3}\langle\!\langle
Y_{-2}|A_+,A_+,A_+\rangle\!\rangle
\right.\\
&~~~~~~~~~\left.+\frac{1}{2}\langle\!\langle
Y_{-2}|A_-,Q_BA_-\rangle\!\rangle
-\langle\!\langle
Y_{-2}|A_+,A_-,A_-\rangle\!\rangle\right].
\end{split}
\label{action7}
\end{equation}
\begin{table}[!h]
\begin{center}
\renewcommand{\arraystretch}{1.4}
\begin{tabular}[h]{||C|c|C|C|C|c||}
\hline \textrm{Notion}& Parity & \textrm{GSO} & \stackrel{\text{\small Superghost}}{\text{\small number}}
 & \textrm{Weight}\, (h)& Comments\\ \hline
\hline A_+ & odd & + & 1 & h\in\Zh,\,h\geqslant -1
& string \\
\cline{1-5} A_- & even & - & 1& h\in\Zh+\frac12,\,h\geqslant -\frac12 & fields\\
\hline \Lambda_+ & even & + & 0& h\in\Zh,\,h\geqslant 0 & gauge\\
\cline{1-5} \Lambda_- & odd & - & 0& h\in\Zh+\frac12,\,h\geqslant \frac{1}{2} & parameters\\ \hline
\end{tabular}
\end{center}
\vspace{-0.5cm}\caption{Parity of string fields and gauge
parameters in the 0 picture.}\label{tab:1}
\end{table}
Here the factors before the odd brackets are fixed by the
constraint of gauge invariance, that is specified below, and
reality of the string fields $A_{\pm}$. Variation of
this action with respect to $A_+$, $A_-$
yields the following equations of motion (see \cite{ABKM} for details)
\begin{equation}
\begin{split}
Q_BA_++A_+\star A_+
-A_-\star A_-&=0,\\
Q_BA_-+A_+\star A_-
-A_-\star A_+&=0.\\
\end{split}
\label{eqmotion}
\end{equation}
The action \eqref{action7} is invariant under the
gauge transformations
\begin{equation}
\begin{split}
\delta A_+&=Q_B\Lambda_++[A_+,\Lambda_+]
+\{A_-,\Lambda_-\},
\\
\delta {A}_-&=Q_B\Lambda_-+[{A}_-,\Lambda_+]
+\{{A}_+,\Lambda_-\},
\end{split}
\label{gauge7}
\end{equation}
where $[\,,]$ ($\{\,,\}$) denotes $\star$-commutator
(-anticommutator)  and $\Lambda_{\pm}$ are gauge parameters (see Table~\ref{tab:1}).

The action \eqref{action7} can be rewritten in the matrix form
 as (see Appendix A)
\begin{equation}
S[\hat{A}]=\frac{1}{2g_o^2}\Tr\left[\frac{1}{2}\int' \hat{A}\star \hat{Q}_B\hat{A}
+\frac13\int'\hat{A}\star\hat{A}\star \hat{A}\right],
\label{theAction}
\end{equation}
where
\begin{equation}
\hat{Q}_B=Q_B\otimes a,\;\;\;\;\;\hat{Y}_{-2}=Y_{-2}\otimes a,
\end{equation}
\begin{equation}
\hat{A}=A_+\otimes a+A_-\otimes b
\label{str:f}
\end{equation}
and $a$ and $b$ are $2\times 2$ matrices such that
\begin{equation}
a^2=1,\quad b^2=-1,\quad\{a,b\}=0.
\end{equation}

One can check the following identity
\begin{equation}
\hat{Q}_B(\hat{\Phi}\star\hat{\Psi})=
(\hat{Q}_B\hat{\Phi})\star\hat{\Psi}+(-1)^{|\hat{\Phi}|}\hat{\Phi}
\star(\hat{Q}_B\hat{\Psi}),
\end{equation}
where $\hat{\Phi}$ and $\hat{\Psi}$ are string fields
\begin{equation}
\hat{\Phi}=\Phi_+\otimes a+\Phi_-\otimes b
\quad\text{and}\quad
\hat{\Psi}=\Psi_+\otimes a+\Psi_-\otimes b
\end{equation}
and parity operator $(-1)^{|\hat{\Phi}|}$ is defined as\footnote{
Our parity operator \eqref{def:parity} can also be
written as $(-1)^{\mathrm{gh}\,\Phi}$ \cite{Ohmori}.
This is true
because all string fields and gauge parameters which we use
satisfy the condition $(-1)^{\mathrm{gh}\,\hat{\Phi}}
(-1)^{|\hat{\Phi}|}=1$.}
\begin{equation}
(-1)^{|\hat{\Phi}|}\Phi_+=(-1)^{|\Phi_+|}\Phi_+
\quad\text{and}\quad
(-1)^{|\hat{\Phi}|}\Phi_-=-(-1)^{|\Phi_-|}\Phi_-.
\label{def:parity}
\end{equation}

The equation of motion following from the action \eqref{theAction}
is
\begin{equation}
\hat{Q}_B\hat{A}+\hat{A}\star\hat{A}=0.
\label{theEquation}
\end{equation}
One can check that the equation \eqref{theEquation} yields
the equations \eqref{eqmotion}. The action \eqref{theAction} is invariant
under the following gauge transformations
\begin{subequations}
\begin{equation}
\delta\hat{A}=\hat{Q}_B\hat{\Lambda}+[\hat{A},\,\hat{\Lambda}],
\label{theGauge}
\end{equation}
where
\begin{equation}
\hat{\Lambda}=\Lambda_+\otimes 1+\Lambda_-\otimes ab.
\label{theLambda}
\end{equation}
\label{GAUGE}
\end{subequations}
It is a matter of simple algebra to check that the
gauge transformations \eqref{theGauge} yield the gauge transformations
\eqref{gauge7} for the fields $A_{\pm}$.\\

\smallskip
\noindent
\textbf{Symmetries}\\
The action \eqref{theAction} is invariant under the following
symmetry transformations:
\begin{subequations}
\begin{enumerate}
\item GSO symmetry. It is given by the following transformations
\begin{equation}
\hat{A}\mapsto ((-1)^F\otimes 1)\hat{A},
\end{equation}
or in components
\begin{equation}
A_{+}\mapsto A_{+}\quad\text{and}\quad A_{-}\mapsto -A_{-}.
\end{equation}
One can also check that the BRST charge $\hat{Q}_B$ commutes with $(-1)^F\otimes 1$;

\item Twist symmetry. The generator of this discrete symmetry
is denoted  by $\Omega$. Its action on the string field is given
by the conformal transformation $M(z)=e^{-\pi i}z$. One can easily
check that the BRST charge $Q_B$ commutes with $\Omega$.
\end{enumerate}
\label{symmetry1}
\end{subequations}

\subsection{Construction of the Cubic Vacuum Superstring Field Theory}
\label{subsec:cubic}
Let $\hat{A}_{0}$ be a solution of the equation \eqref{theEquation}.
A shift of a string field $\hat{A}$
\begin{equation*}
\hat{A}=\hat{A}_0+\hat{\Ac}.
\end{equation*}
yields the following form of the action
\eqref{theAction}
\begin{equation}
S[\hat{A}_0,\hat{\Ac}]=S[\hat{A}_0]+\frac{1}{2g_o^2}\Tr\left[\frac{1}{2}\int' \hat{\Ac}\star \hat{Q}\hat{\Ac}
+\frac13\int'\hat{\Ac}\star\hat{\Ac}\star \hat{\Ac}\right],
\label{VSFTaction}
\end{equation}
where $\hat{Q}$ is ``a new BRST charge'' of the form
\begin{equation}
\hat{Q}=\hat{Q}_B+\{\hat{A_0},\cdot\}.
\label{shift:BRST}
\end{equation}
Further we will refer to $\hat{Q}$ as to kinetic operator.
One can check that the equation $\hat{Q}^2=0$
yields the equation of motion for the field $\hat{A_{0}}$ and
therefore $\hat{Q}$ is nilpotent.

Now let us investigate the structure of the shifted BRST charge more carefully.
Consider an arbitrary string field $\hat{\Phi}$, then
the operator \eqref{shift:BRST} acts on it as follows
\begin{multline}
\hat{Q}\hat{\Phi}=Q_B\Phi_+\otimes 1+Q_B\Phi_-\otimes ab
+\{A_{0,+}\otimes a+A_{0,-}\otimes b,\Phi_+\otimes a+\Phi_-\otimes b\}
\\
=\left(Q_B\Phi_+ + \{A_{0,+},\Phi_+\}-\{A_{0,-},\Phi_-\}\right)\otimes 1
+(Q_B\Phi_-+[A_{0,+},\Phi_-]-[A_{0,-},\Phi_+])\otimes ab.
\label{qeq1}
\end{multline}
Let us introduce two new operators $Q_{\textsf{odd}}$ and $Q_{\textsf{even}}$
\footnote{The generalization of \eqref{shift:BRST} and \eqref{abg:1}
for the case of the arbitrary ghost number of $Z$ was given
in \cite{Ohmori}:
\begin{equation}
\hat{Q}\hat{Z}=\hat{Q}_B\hat{Z}+\hat{A_0}\star\hat{Z}-(-1)^{\text{gh}(\hat{Z})}\hat{Z}\star\hat{A_0}
\label{shift:BRST-general}
\end{equation}
and
\begin{subequations}
\begin{align}
Q_{\textsf{odd}}Z&=Q_BZ+A_{0,+}\star Z-(-1)^{|Z|}Z\star A_{0,+},
\\
Q_{\textsf{even}}Z&=A_{0,-}\star Z-(-1)^{\text{GSO}(Z)}Z\star A_{0,-}.
\label{abg:1-general}
\end{align}
\label{abg:44-general}
\end{subequations}
This coincides with \eqref{shift:BRST} and \eqref{abg:1} for odd ghost number.}:
\begin{subequations}
\begin{align}
Q_{\textsf{odd}}Z&=Q_BZ+A_{0,+}\star Z-(-1)^{|Z|}Z\star A_{0,+},
\\
Q_{\textsf{even}}Z&=A_{0,-}\star Z+(-1)^{|Z|}Z\star A_{0,-},
\label{abg:1}
\end{align}
\label{abg:44}
\end{subequations}
where $Z$ is a string field in GSO$+$ or GSO$-$ sector and $|Z|$ is a parity
of the field $Z$.
One sees that \eqref{qeq1} means that $\hat{Q}$
can be written in the form
\begin{equation}
\hat{Q}=Q_{\textsf{odd}}\otimes a+Q_{\textsf{even}}\otimes b.
\label{abg:100}
\end{equation}
The nilpotency of the $\hat{Q}$
yields the following identities for the operators $Q_{\textsf{odd}}$
and $Q_{\textsf{even}}$
\begin{equation}
Q_{\textsf{odd}}^2-Q_{\textsf{even}}^2=0\quad\text{and}\quad
[Q_{\textsf{odd}},\,Q_{\textsf{even}}]=0.
\label{eqBRST}
\end{equation}\\
\textbf{Symmetries}\\
Let us now discuss what happened to the symmetries \eqref{symmetry1}.
\begin{enumerate}
\label{symm2}
\item GSO symmetry. It is now broken. This happens
because the generator of this symmetry $(-1)^F$ does not
commute this the kinetic operator \eqref{shift:BRST}.
Under the action of $(-1)^F$ the kinetic operator
\eqref{abg:44}, \eqref{abg:100} transforms to
\begin{equation*}
Q_{\textsf{odd}}\mapsto Q_{\textsf{odd}}\quad
\text{and}\quad
Q_{\textsf{even}}\mapsto -Q_{\textsf{even}}.
\end{equation*}
It is very natural that GSO symmetry is broken when
we consider a theory around the vacuum solution.
The reason is the following \cite{ZWp}: the original theory
has GSO symmetry, therefore if $(A_{0,+},A_{0,-})$
is a solution of the EOM, then so is $(A_{0,+},-A_{0,-})$
(see Figure~\ref{fig:1}).
And therefore we have to have \textit{two}
kinetic operators corresponding to two possible vacuum solutions.
This kinetic operators are related to each other by GSO transformation.

\item Twist symmetry. In general
the twist symmetry is broken. The only
class of the kinetic operators which allows unbroken twist symmetry
are the ones corresponding to the solutions satisfying
$\Omega(A_{0,+})=A_{0,+}$
and $\Omega(A_{0,-})=A_{0,-}$.
\end{enumerate}

\subsection{VSSFT Equations of Motion}
Equations of motion following from the VSFT action \eqref{VSFTaction}
have the same form as \eqref{theEquation} but with the shifted BRST operator $\hat{Q}$.
In components these equations are
\begin{equation}
\begin{split}
&Q_{\textsf{odd}}\Ac_+-Q_{\textsf{even}}\Ac_-+\Ac_+\star \Ac_+
-\Ac_-\star \Ac_-=0,\\
&Q_{\textsf{odd}}\Ac_--Q_{\textsf{even}}\Ac_+
+\Ac_+\star\Ac_--\Ac_-\star\Ac_+=0.\\
\end{split}
\label{eqmotion-cal'}
\end{equation}
It is more convenient to rewrite these equations in ``light-cone''
variables $\Ac$ and $\bar{\Ac}$:
\begin{equation}
\Ac=\Ac_+-\Ac_-\quad\text{and}\quad
\bar{\Ac}=\Ac_++\Ac_-.
\label{lighcone-strfield}
\end{equation}
Then equations \eqref{eqmotion-cal'} have the following simple form:
\begin{equation}
q\bar{\Ac}+\Ac\star\bar{\Ac}=0
\quad
\text{and}
\quad
\bar{q}\Ac+\bar{\Ac}\star\Ac=0,
\label{EOM}
\end{equation}
where
\begin{equation}
q=Q_{\textsf{odd}}-Q_{\textsf{even}}
\quad\text{and}\quad
\bar{q}=Q_{\textsf{odd}}+Q_{\textsf{even}}.
\label{q-q}
\end{equation}
From the relations \eqref{eqBRST} one can get the following
properties of the charges $q$ and $\bar{q}$:
\begin{equation}
q\bar{q}=0\quad\text{and}\quad
\bar{q}q=0.
\label{q-properties}
\end{equation}
It is also fruitful to rewrite the gauge transformation
\begin{equation*}
\delta\hat{\Ac}=\hat{Q}\hat{\Lambda}+[\hat{\Ac},\,\hat{\Lambda}]
\end{equation*}
in the light-cone variables:
\begin{equation}
\Lambda=\Lambda_+-\Lambda_-
\quad\text{and}\quad
\bar{\Lambda}=\Lambda_++\Lambda_-.
\label{lcGauge}
\end{equation}
The gauge transformations become
\begin{equation}
\begin{split}
\delta \Ac&=q\Lambda+A\star\Lambda-\bar{\Lambda}\star A,
\\
\delta \bar{\Ac}&=\bar{q}\bar{\Lambda}+\bar{A}\star\bar{\Lambda}-\Lambda\star\bar{A}.
\end{split}
\end{equation}


\section{Construction of the Ghost Kinetic Operators}
\label{sec:newBRST}
\subsection{Restrictions on $\hat{Q}$ following from Sen conjectures
}
According to Sen conjectures \cite{sen-con} the solution $\hat{A}_0$
represents the
vacuum without open string excitations\footnote{This conjecture has been checked
for the non-BPS brane decay only at the first non-trivial
level {\cite{ABGKM}}.}, and therefore the cohomology of the kinetic
operator $\hat{Q}$ must be zero.

As in the bosonic case \cite{F2}
here it also might be easier to guess the form of the BRST charge then
to derive it.
In proposing a simple form of the vacuum SSFT action, we have in mind
field redefinition, which preserves the form of the cubic action,
but simplifies the expression for the kinetic operator $\hat{Q}$.
By an appropriate field redefinition
\begin{subequations}
\begin{equation}
\hat{\Uc}=\Uc_{\textsf{even}}\otimes 1+\Uc_{\textsf{odd}}\otimes ab
\end{equation}
we will assume a $\star$-algebra homomorphism
\begin{equation}
\hat{\Uc}(\hat{\Ac}\star\hat{\Bc})=(\hat{\Uc}\hat{\Ac})\star(\hat{\Uc}\hat{\Bc}),
\end{equation}
which satisfy two additional conditions: the invariance of
the integral with respect to this homomorphism
\begin{equation}
\Tr\int'\hat{\Uc}\hat{\Ac}=\Tr\int'\hat{\Ac};
\end{equation}
and the existence of the right inverse
\begin{equation}
\hat{\Uc}\hat{\Uc}^{-1}=1.
\end{equation}
\label{def:redef}
\end{subequations}
The  $\hat{}$  in the expressions for the field redefinition $\hat{\Uc}$
is very important since this transformation acts in both GSO$+$ and GSO$-$ sectors.
Using \eqref{def:redef} one can check that after the field redefinition
\begin{equation*}
\hat{\Ac}\mapsto \hat{\Uc}\hat{\Ac}
\end{equation*}
the kinetic operator transforms into
\begin{equation}
\hat{\Qc}=\hat{\Uc}^{-1}\hat{Q}\hat{\Uc}.
\end{equation}
Note that the transformation $\hat{\Uc}$ is highly
non-trivial and mixes GSO$+$ and GSO$-$
sectors.

Now it can be useful to consider an example of
the field redefinition that can seriously simplify an expression
for BRST charge. Let consider the standard BRST charge in the
superconformal field theory
\begin{equation}
Q_B=\frac{1}{2\pi i}\oint d\zeta \Bigl[
c(T_B+T_{\phi}+T_{\eta\xi}+\frac{1}{2}T_{bc})
-\eta e^{\phi}T_F+\frac{1}{4}b\pd\eta\eta e^{2\phi}
\Bigr].
\label{BRST:std}
\end{equation}
One can check that after the homogenous field redefinition
\cite{9902178}
\begin{equation}
\Uc=e^{-R},\quad\text{where}\quad
R=\frac{1}{2\pi i}\oint d\zeta \Bigl[
cT_F e^{-\phi}e^{\chi}+\frac{1}{4}\pd(e^{-2\phi})e^{2\chi}c\pd c
\Bigr]\label{redef-1}
\end{equation}
the BRST charge \eqref{BRST:std} takes the form
\begin{equation}
\Qc=\Uc^{-1}Q_B\Uc=\frac{1}{2\pi i}\oint d\zeta\,
b\gamma^2(\zeta).\label{redef-2}
\end{equation}
\\

Following the idea of the paper \cite{F2},
which is based on Sen conjectures, gauge invariance
and algebraic properties of the BRST charge,
we require $\hat{\Qc}$ to satisfy the following properties:
\begin{subequations}
\begin{enumerate}
\item $\hat{\Qc}=\Qc_{\textsf{odd}}\otimes a + \Qc_{\textsf{even}}\otimes b$;
\item Both $\Qc_{\textsf{odd}}$ and $\Qc_{\textsf{even}}$ have superghost number
equal to one, but $\Qc_{\textsf{odd}}$ is Grassmann odd, while
$\Qc_{\textsf{even}}$ is Grassmann even;
\item $\hat{\Qc}$ is a nilpotent operator, that in components means
the identities
\begin{equation}
\Qc_{\textsf{odd}}^2-\Qc_{\textsf{even}}^2=0
\quad\text{and}\quad
[\Qc_{\textsf{odd}},\Qc_{\textsf{even}}]=0;
\label{Qequations}
\end{equation}
\item $\hat{\Qc}$ is a differentiation of the $\star$-algebra
\begin{equation}
\hat{\Qc}(\hat{A}\star\hat{B})=(\hat{\Qc}\hat{A})\star\hat{B}+(-1)^{|\hat{A}|}
\hat{A}\star(\hat{\Qc}\hat{B}),
\end{equation}
where the parity operator $(-1)^{|\hat{A}|}$ was defined in \eqref{def:parity}.
In particular, this identity means that operators $\Qc_{\textsf{odd}}$ and $\Qc_{\textsf{even}}$
also satisfy the Leibnitz rule;
\item The integral of the full derivative is zero
\begin{equation}
\Tr \int'\hat{\Qc}(\hat{\Ac}\star\hat{\Bc})=0;
\end{equation}
\item The operator $\hat{\Qc}$ must be universal, what means
that it has to be written without reference to the brane
boundary CFT;
\item The operator $\hat{\Qc}$ must have vanishing cohomology;
\item The operator $\hat{\Qc}$ must commute with the double step inverse picture-changing
operator
\begin{equation}
[\hat{Y}_{-2},\,\hat{\Qc}]=0\quad\text{or}\quad
\{\hat{Y}_{-2},\,\hat{\Qc}\}=0.
\end{equation}
We need this axiom to relate the axiom~5 with the fact
that $\hat{\Qc}$ annihilates the identity $|\Ic\rangle$. Therefore
we can have several variations of this axiom and in general
we only need something like the following
\begin{equation*}
\Qc_{\textsf{odd}}Y_{-2}\pm Y_{-2}\Qc_{\textsf{odd}}=0\quad\text{and}\quad
\Qc_{\textsf{even}}Y_{-2}\pm Y_{-2}\Qc_{\textsf{even}}=0;
\end{equation*}
Plus/minus in these formulae can be chosen independently.
\item $\hat{\Qc}$ is a hermitian operator, which means that both
$Q_{\textsf{odd}}$ and $Q_{\textsf{even}}$ are hermitian ones.
\end{enumerate}
We will construct the operator satisfying these requirements
in the next subsection.
\label{BRST:cond}
\end{subequations}

Note that in the case of non-polynomial superstring
field theory $\cite{0002211}$ all the requirements remains the same, but the
axiom~8 about commutation
with the double step inverse picture-changing operator
has to be changed into requirement of anticommutation
with $\hat{\eta}_0$ (see Appendix~\ref{app:B} for details).

\subsection{Construction of the new kinetic operator}
The simplest way to satisfy the conditions \eqref{BRST:cond}
is to put $\Qc_{\textsf{even}}=0$ and choose
$\Qc_{\textsf{odd}}$ as in the bosonic theory or put
\begin{equation*}
\Qc_{\textsf{odd}}=\frac{1}{2\pi i}\oint d\zeta\, b\gamma^2(\zeta).
\end{equation*}
However since $A_{0,+}\neq 0$ and $A_{0,-}\neq 0$ we believe that after the field
redefinition both charges $\Qc_{\textsf{odd}}$ and
$\Qc_{\textsf{even}}$ are non zero.

One can try to take the following formal expression for the ghost
kinetic operator \footnote{We have discussed this form of the
ghost kinetic operator in the first version of this paper}
\begin{subequations}
\begin{align}
\Qc_{\textsf{odd}}&=\mu^2\,c(i)+\frac{1}{2\pi i}\oint
b(z)\gamma^2(z)dz,
\\
\Qc_{\textsf{even}}&=\mu\,\gamma(i),
\end{align}
\label{formal-q-old}
\end{subequations}
where $\mu$ is a complex number. This kinetic operator satisfies
the conditions 1-8, however, as it has been noted by Ohmori
\cite{Ohmori} this operator does not satisfy 9. The following
modification of (\ref{formal-q-old}) has been proposed in
\cite{Ohmori}
\begin{subequations}
\begin{align}
\Qc_{\textsf{odd}}&=\frac{\mu^2}{4i}\,\bigl[c(i)-c(-i)\bigr]
+\frac{1}{2\pi i}\oint b(z)\gamma^2(z)dz,
\\
\Qc^{+}_{\textsf{even}}&=\frac{\mu}{2i}\,\bigl[\gamma(i)-\gamma(-i)\bigr],\\
\Qc^{-}_{\textsf{even}}&=\frac{\mu}{2}\,\bigl[\gamma(i)+\gamma(-i)\bigr],
\end{align}
\label{formal-q}
\end{subequations}
where  $\Qc^{\pm}_{\textsf{even}}$ means the restriction of the
operator $\Qc_{\textsf{even}}$ to GSO$\pm$ sectors. In some sense
\eqref{formal-q} is the only form for the kinetic operator which
satisfies the twist invariance and the conditions
\eqref{BRST:cond}. One can explain it as follows. Following
\cite{rsz-6} consider an original (before field redefinition) BRST
charge $Q$ defined as
\begin{equation}
Q=\sum_{r}\frac{1}{2\pi i}\oint d\zeta\, a_{r}(\zeta)\Oc_{r}(\zeta)
\label{ser}
\end{equation}
where $a_{r}$ are smooth forms of $\zeta$ and $\Oc_{r}(\zeta)$ are some local
conformal operators of ghost number $1$. It was shown \cite{rsz-6} that
after a singular field redefinition
the  dominant contribution to the transformed charge $\Qc$
will come from the lowest dimensional conformal operators.
This has led to the choice of $c(i)$ and $c(-i)$ in the bosonic case,
and this also leads to our choice of $\Qc_{\textsf{even}}$, since $\gamma$ is
the lowest dimensional even primary operator of ghost number $1$.

The ansatz \eqref{formal-q} obviously satisfies the axiom 2 from the
previous subsection. The axiom 3 about
nilpotency of $\hat{\Qc}$
is also satisfied, because
\begin{subequations}
\begin{equation}
\Qc_{\textsf{odd}}^{2}\equiv\Qc^{\mp}_{\textsf{even}}\Qc^{\pm}_{\textsf{even}}
=\frac{\mu^2}{4i}(\gamma^{2}(i)-\gamma^{2}(-i))
\end{equation}
and since there is no $\beta$ in the expression for $\Qc_{\textsf{odd}}$ one gets
\begin{equation}
[\Qc_{\textsf{odd}},\,\Qc_{\textsf{even}}]=0.
\end{equation}
\end{subequations}

The most non trivial is to check the axiom 4, which in particular says
that $\Qc_{\textsf{odd}}$ and $\Qc_{\textsf{even}}$ admit the Leibnitz
rule and the axiom 5, which says that integral of ``the full
derivative'' is zero. These axioms are also related to the hermitian property
of the kinetic ghost operator
\begin{equation}
\Tr\, \la\hat{\Ac},\,\hat{\Qc}\hat{\Bc}\ra
=\Tr\, \la\hat{\Qc}\hat{\Ac},\,\hat{\Bc}\ra
\label{herm-prop}
\end{equation}
and the gauge invariance
of the vacuum string field theory action. It is a matter of
overlap equations that if we could proof the axiom~5, then
the axiom~4 will be automatically satisfied.

\subsubsection{Check of axiom~5.}
It doesn't seem that $\Qc_{\textsf{odd}}$ and $\Qc_{\textsf{even}}$,
which contain midpoint
insertions, can satisfy axiom~5. The reason is that they
diverge acting on the identity $|\Ic\rangle$.
However, one can define
$\Qc_{\textsf{odd}}$ and $\Qc_{\textsf{even}}$ as a limit of a sequence
with each element annihilating the identity.

To construct such a sequence let us consider the overlap equation
for the identity. To write it one has to be very careful, because
in the superconformal CFT we have conformal fields with half integer weights.
We will also assume that we make double trick for antiholomorphic operators,
and therefore the overlap equation will connect the fields
on the boundary of the unit disk, and not only on the boundary of the
upper half unit disk. The argument of coordinate $z$ on the
unit disk is in interval $(-\pi,\pi)$.
The following
overlap equations for a conformal operator $\Oc_h$ of the weight $h$ take place:
\begin{subequations}
\begin{align}
\left[\Oc_{h}(z)
-\left(\frac{1}{z^{2}}\right)^{h}\Oc_{h}\left(\frac{e^{i\pi}}{z}\right)
\right]
|\Ic\rangle&=0,\quad\text{for}\quad |z|=1,\,\Re z<0\quad\text{and}
\quad \Im z>0;
\\
\left[\Oc_{h}(z)
-\left(\frac{e^{-2\pi i}}{z^{2}}\right)^{h}\Oc_{h}\left(\frac{e^{-i\pi}}{z}\right)
\right]
|\Ic\rangle&=0,\quad\text{for}\quad |z|=1,\,\Re z<0\quad\text{and}
\quad \Im z<0.
\end{align}
\end{subequations}
So one sees that if $h$ is a half integer the second equation
differs from the first by a sign of the second term.

Now the regularization is clear. Let us simply define kinetic operators
$\Qc^{\eps}_{\textsf{odd}}$ and $\Qc^{\eps}_{\textsf{even}}$ by substitutions
\begin{subequations}
\begin{align}
c(i)&\mapsto \frac{1}{2}\bigl[e^{-i\eps}c(ie^{i\eps})+e^{i\eps}c(ie^{-i\eps})
\bigr],\\
\gamma(i)&\mapsto \frac{1}{e^{-i\pi/4}-e^{i\pi/4}}
\bigl[e^{-i\pi/4-i\eps/2}\gamma(ie^{i\eps})
-e^{i\pi/4+i\eps/2}\gamma(ie^{-i\eps})\bigr].
\end{align}
\label{I:overlap}
\end{subequations}
This regularization corresponds to a splitting of the midpoint
insertion into two
insertions on the left-half and on the right-half of the string.
In limit $\eps\rightarrow 0$ one recovers the midpoint insertion.
From overlap equations \eqref{I:overlap} it follows that
\begin{subequations}
\begin{align}
\Bigl[e^{-i\eps}c(ie^{i\eps})+e^{i\eps}c(ie^{-i\eps})\Bigr]
|\Ic\rangle&=0,
\\
\Bigl[e^{-i\pi/4-i\eps/2}\gamma(ie^{i\eps})
-e^{i\pi/4+i\eps/2}\gamma(ie^{-i\eps})\Bigr]
|\Ic\rangle&=0.
\end{align}
\end{subequations}
One can now define $\Qc_{\textsf{odd}}$ and $\Qc_{\textsf{even}}$ as
\begin{gather}
\Qc_{\textsf{odd}}\equiv \lim_{\eps\rightarrow0}\Qc^{\eps}_{\textsf{odd}}\quad\text{and}\quad
\Qc_{\textsf{even}}\equiv \lim_{\eps\rightarrow0}\Qc^{\eps}_{\textsf{even}}.
\end{gather}
This finishes the proof that $\Qc_{\textsf{odd}}$ and $\Qc_{\textsf{even}}$
annihilate the identity.

\subsubsection{Check of axiom~7. Zero cohomology.}

As in the bosonic case \cite{rsz-6} the equation
$\hat{\Qc}\hat{\Phi}=0$ has no non-zero solutions which belong to Fock space
\footnote{This is so because $\hat{\Qc}$
involves oscillators of all possible levels, while states in Fock space
have to be polynomials.}.
Let us suppose that there
is a generalized state $\hat{\Psi}$ annihilated by $\hat{\Qc}$.
We want to show that for any such $\hat{\Psi}$
there exists $\hat{\Lambda}$
such that
\begin{equation}
\hat{\Psi}=\hat{\Qc}\hat{\Lambda}.
\label{cohom:1}
\end{equation}
To show this it is sufficient to find an operator $\hat{\Kc}$
such that
\begin{equation}
\{\hat{\Qc},\,\hat{\Kc}\}=\mathrm{Id}\otimes 1.
\label{cohom:2}
\end{equation}
Indeed, acting by $\hat{\Kc}\hat{\Qc}$ onto the expression \eqref{cohom:1}
\begin{equation*}
0=\hat{\Kc}(\hat{\Qc}\hat{\Psi})\stackrel{\ref{cohom:2}}{=}\hat{\Psi}
-\hat{\Qc}(\hat{\Kc}\hat{\Psi}),
\end{equation*}
we find  $\hat{\Lambda}=\hat{\Kc}\hat{\Psi}$.

It is natural   to search for the operator $\hat{\Kc}$ in the following form
\begin{gather}
\hat{\Kc}=\Kc_{+}\otimes a+\Kc_{-}\otimes b.
\end{gather}
The equation \eqref{cohom:2} yields
\begin{equation*}
\mathrm{Id}\otimes 1=
\{\hat{\Qc},\hat{\Kc}\}=\bigl(\{\Qc_{\textsf{odd}},\Kc_{+}\}
-\{\Qc_{\textsf{even}},\Kc_{-}\}\bigr)\otimes 1
+\bigl(-[\Qc_{\textsf{even}},\Kc_{+}]+[\Qc_{\textsf{odd}},\Kc_{-}]\bigr)\otimes ab.
\end{equation*}
For the operators $\Qc_{\textsf{odd}}$ and $\Qc_{\textsf{even}}$
defined by \eqref{formal-q} one can take
\begin{gather}
\Kc_-=0\quad\text{and}\quad \Kc_+=\frac{2}{\mu^2}\,b_0.
\end{gather}

\subsubsection{Check of axiom~8.}
We show that the following relations are true
\begin{gather}
[\Qc_{\textsf{odd}},Y_{-2}]=0\quad\text{and}\quad
[\Qc_{\textsf{even}},Y_{-2}]=0.
\end{gather}
The first commutator is obviously zero,
since $Y_{-2}$ commutes with $c(\cdot)$. It also commutes
with $\oint\, b\gamma^2$, since it is a part of the original BRST charge.
Let us check the second commutator. To this end let us remind, that
\begin{equation*}
Y_{-2}=Y(i)Y(-i),\quad Y(z)=4c\pd\xi e^{-2\phi}(z)
\quad\text{and}\quad \gamma(z)=\eta e^{\phi}(z).
\end{equation*}
Consider the following OPEs:
\begin{align*}
Y(i)Y(-i)\gamma(z)&=
-c(i)c(-i)\Bigl[
:\pd\xi(i)\pd\xi(-i)\eta(z):-\frac{\pd\xi(i)}{(i+z)^2}
+\frac{\pd\xi(-i)}{(i-z)^2}
\Bigr](z^2+1)^2e^{-2\phi(i)-2\phi(-i)+\phi(z)}
\\
\gamma(z)Y(i)Y(-i)&=
-c(i)c(-i)\Bigl[
:\pd\xi(i)\pd\xi(-i)\eta(z):-\frac{\pd\xi(i)}{(i+z)^2}
+\frac{\pd\xi(-i)}{(i-z)^2}
\Bigr](z^2+1)^2e^{-2\phi(i)-2\phi(-i)+\phi(z)}
\end{align*}
So one sees that $[\gamma(z),Y_{-2}]=0$ for any $z$. Now assuming
that we are dealing with the regularized charge $\Qc^{\eps}_{\textsf{even}}$
we get that it commutes with the double step inverse insertion operator.
And after taking the limit $\eps\rightarrow 0$ we obtain
the result required for $\Qc_{\textsf{even}}$.

\subsubsection{Check of axiom 9.}
It has been shown in \cite{Ohmori} that the hermitian property
\eqref{herm-prop} holds  due to the following conformal properties
of the kinetic operator
\begin{subequations}
\begin{gather}
\Qc^{+}_{\textsf{even}}=-I\circ \Qc^{-}_{\textsf{even}},\\
\Qc^{-}_{\textsf{even}}=I\circ \Qc^{+}_{\textsf{even}},
\end{gather}
\end{subequations}
where $I(z)=-1/z$.

\subsubsection{Symmetries.}
Let us now check that the ghost kinetic operators (\ref{formal-q})
 satisfy the symmetries we have discussed on page
\pageref{symm2}.
\begin{enumerate}
\item GSO symmetry. GSO transformation
relates the ghost kinetic operators \eqref{formal-q} with $\mu$
and $-\mu$.

\item Twist symmetry.
Twist transformation $\Omega$ changes $i\mapsto -i$ and
\begin{subequations}
\begin{gather}
\Omega c(i)\Omega^{-1}=-\, c(-i),~~~
\Omega c(-i)\Omega^{-1}=-\,
c(i),\\
\Omega \gamma(i)\Omega^{-1}=i\,\gamma(-i),~~~
\Omega
\gamma(-i)\Omega^{-1}=i\,\gamma(i).
\end{gather}
\end{subequations}
The twist operator
$\Omega$ acts in the following way on the string
fields of definite weight  \cite{ABKM}
\begin{subequations}
\begin{gather}
\Omega (|\Ac_{+}\rangle)=(-1)^{h_{\Ac_+}+1}|\Ac_{+}\rangle,\\
\Omega (|\Ac_{-}\rangle)=(-1)^{h_{\Ac_-}+\frac12}|\Ac_{-}\rangle.
\end{gather}
\end{subequations}
Using this fact Ohmori \cite{Ohmori} has shown  that the following
identities hold
\begin{subequations}
\begin{gather}
\Omega(\Qc^{+}_{\textsf{even}}|\Ac_{+}\rangle)=\Qc^{+}_{\textsf{even}}\Omega(|\Ac_{+}\rangle),\\
\Omega(\Qc^{-}_{\textsf{even}}|\Ac_{-}\rangle)=\Qc^{-}_{\textsf{even}}\Omega(|\Ac_{-}\rangle).
\end{gather}
\end{subequations}
This means that the twist
invariant choice of the kinetic operator $\Qc_{\textsf{even}}$
takes the different form in GSO$+$ and GSO$-$ sectors.
\end{enumerate}

\section{Conclusion and Discussions}
In this paper we have considered a simplest candidate
for the kinetic operator of the cubic VSSFT which describes a result
of a decay of unstable non-BPS brane.
In spite of the fact that we consider only cubic superstring field
theory, our results concerning the ghost kinetic operator
can be applied to the Berkovits non-polynomial superstring field theory
(see Appendix~\ref{app:B}).

As the problems to be solved let us note
\begin{itemize}
\item Find the solution of the ghost equations of motion\footnote{Solutions of the ghost equations of motion has been
discussed recently in \cite{AGK},\cite{Ohmori}};
\item Question about the restoration of the supersymmetry in
the non-perturbative vacuum.
\end{itemize}
\section*{Acknowledgments}

We would like to thank
 A.S.Koshelev for discussions.
 This work was supported in part by RFBR grant
02-01-00695  and RFBR grant for leading scientific schools and  by
INTAS grant 99-0590.

\appendix
\section*{Appendix}
\addcontentsline{toc}{section}{Appendix}
\renewcommand {\theequation}{\thesection.\arabic{equation}}

\section{Matrix Representation of non-GSO Projected String Fields}\label{app:A}
\setcounter{equation}{0}
The cubic action involving these two sectors was constructed
by the authors in \cite{ABKM}. Here we will rewrite it
in more simple form using only one (matrix valued) string field:
\begin{equation}
\hat{\Ac}=\Ac_+\otimes a+\Ac_-\otimes b,
\label{str1}
\end{equation}
where $a$ and $b$ are $N\times N$ matrices we wish to find.
We have to write also
\begin{equation}
\hat{Q}_B=Q_B\otimes q\qquad\text{and}\qquad
\hat{Y}_{-2}=Y_{-2}\otimes y,
\end{equation}
where $q$ and $y$ are also $N\times N$ matrices we need to determine.
We propose the action for string field \eqref{str1} in the form
\begin{equation}
S=\frac{1}{g_o^2}\frac{1}{N}\Tr\left[\frac{1}{2}\int' \hat{\Ac}\star \hat{Q}_B\hat{\Ac}
+\frac13\int'\hat{\Ac}\star\hat{\Ac}\star \hat{\Ac}\right],
\end{equation}
where $\Tr$ is trace over the matrix multiplier and $\int'$ denotes
Witten's integration but with the insertion of double step
inverse picture changing operator $Y_{-2}$, i.e.  $\int'=\int \hat{Y}_{-2}$.
One can check that to obtain the action proposed in \cite{ABKM}
the matrices $q,\,y,\,a,\,b$ have to satisfy the following conditions:
\begin{subequations}
\begin{align}
\Tr(yaqa)&=\Tr(ybqb)=N;
\label{c1}
\\
\Tr(ya^3)&=N;
\label{c2}
\\
\Tr(yab^2)&=-\Tr(ybab)=\Tr(yb^2a);
\label{c3}
\\
\Tr(yab^2)&=-N.
\label{c4}
\end{align}
\end{subequations}
The condition \eqref{c3} is satisfied if
\begin{subequations}
\begin{equation}
[a,b^2]=0\quad\text{and}\quad \{a,b\}=0.
\end{equation}
The condition \eqref{c2} is satisfied if
\begin{equation}
a^2=1\quad\text{and}\quad y=a.
\end{equation}
And the conditions \eqref{c1} and \eqref{c4} are satisfied if
\begin{equation}
q=a\quad\text{and}\quad b^2=-1.
\end{equation}
\label{alg1}
\end{subequations}
So we are left with two matrices $a$ and $b$ such that
\begin{equation}
a^2=1,\quad b^2=-1\quad\text{and}\quad \{a,b\}=0.
\end{equation}
Further we will assume that $a$ and $b$ are $2\times 2$ matrices,
for example
\begin{equation}
a=\sigma_3,\quad b=i\sigma_2 \quad\text{and}\quad ab=\sigma_1.
\label{sigma}
\end{equation}

\section{Application to non-polynomial String Field Theory}\label{app:B}
\setcounter{equation}{0}
Let us remind the action for the NS sector of the non-polynomial super string field theory
on a non-BPS brane \cite{0002211}:
\begin{equation}
S[\hat{\Phi}]=\frac{1}{4}\Tr\int\left[
(e^{-\hat{\Phi}}\hat{Q}_Be^{\hat{\Phi}})(e^{-\hat{\Phi}}\hat{\eta}_0 e^{\hat{\Phi}})
-\int_0^1dt\,
(e^{-t\hat{\Phi}}\pd_t e^{t\hat{\Phi}})
\{(e^{-t\hat{\Phi}}\hat{Q}_B e^{t\hat{\Phi}}),\,
(e^{-t\hat{\Phi}}\hat{\eta}_0 e^{t\hat{\Phi}})\}\right],
\label{ber}
\end{equation}
where
\begin{equation}
\hat{\Phi}=\Phi_+\otimes 1+\Phi_-\otimes ab,
\quad \hat{Q}_B=Q_B\otimes a\quad\text{and}\quad
\hat{\eta}_0=\eta_0\otimes a.
\label{str:b}
\end{equation}
Here matrices $a$ and $b$ are defined by \eqref{sigma}, $\Phi_+$ and $\Phi_-$
are picture zero superghost zero string fields
in GSO$+$ (Grassmann even) and GSO$-$ (Grassmann odd) sector respectively

It is shown in the papers \cite{0105319} and \cite{Marino} that the
action for the shifted string field $\hat{h}$ has the same form
as the original one \eqref{ber}, but with the shifted BRST charge
$\hat{Q}$, which is defined on an arbitrary string field $\hat{\Phi}$ as
\begin{equation}
\hat{Q}\hat{\Phi}=\hat{Q}_B\hat{\Phi}+[\hat{A},\,\hat{\Phi}].
\label{ber:BRST}
\end{equation}
Here
\begin{equation}
\hat{A}=e^{-\hat{\Phi}_0}\hat{Q}_B e^{\hat{\Phi}_0}
\end{equation}
and $\Phi_0$ is a solution of the equations of motion following
from \eqref{ber}. Using the matrix representations for $\hat{Q}_B$
and $\hat{\Phi}_0$ one can show that $\hat{A}$ can be represented
in the following form
\begin{equation}
\hat{A}=A_{\textsf{odd}}\otimes a+A_{\textsf{even}}\otimes b.
\end{equation}
Now let us substitute this representation into \eqref{ber:BRST}
\begin{multline}
\hat{Q}\hat{\Phi}=Q_B\Phi_+\otimes a+Q_B\Phi_-\otimes b
+[A_{\textsf{odd}}\otimes a+A_{\textsf{even}}\otimes b,\,\Phi_+\otimes 1+
\Phi_-\otimes ab]
\\
=(Q_B\Phi_++[A_{\textsf{odd}},\,\Phi_+]+\{A_{\textsf{even}},\,\Phi_-\})\otimes a
+(Q_B\Phi_-+\{A_{\textsf{odd}},\,\Phi_-\}+[A_{\textsf{even}},\,\Phi_+])\otimes b.
\label{ber:77}
\end{multline}
Let us introduce two new operators $Q_{\textsf{odd}}$ and $Q_{\textsf{even}}$:
\begin{subequations}
\begin{align}
Q_{\textsf{odd}}X&=Q_BX+A_{\textsf{odd}}\star X-(-1)^{|X|}X\star A_{\textsf{odd}},
\\
Q_{\textsf{even}}X&=A_{\textsf{even}}\star X-(-1)^{|X|}X\star A_{\textsf{even}},
\label{ber:1}
\end{align}
\label{ber:44}
\end{subequations}
where $X$ is a string field in GSO$+$ or GSO$-$ sector and $|X|$ is a parity
of the field $X$. Notice that the difference between the formulae \eqref{abg:1}
and \eqref{ber:1} in the sign is only due to the difference in the definition
of the string field (compare \eqref{str:f} and \eqref{str:b}), but
in principle the operators defined by \eqref{ber:44} and \eqref{abg:44}
are the same.
One sees that \eqref{ber:77} means that $\hat{Q}$
can be written in the form
\begin{equation}
\hat{Q}=Q_{\textsf{odd}}\otimes a+Q_{\textsf{even}}\otimes b.
\label{ber:55}
\end{equation}
The nilpotency\footnote{Let us notice the difference in cubic and non-polynomial
SFTs. In the cubic SFT the nilpotency of the shifted BRST charge yields
the equations of motion for the background field, while in the non-polynomial
one the nilpotency of the shifted BRST charge is just an algebraic identity.
The equations of motion for the background field are obtained
from the requirement $\{\eta_0,\,Q\}=0$. Therefore one has
to add this condition to the axioms in the Section~\ref{sec:newBRST}.} of the $\hat{Q}$
yields the following identities for the operators $Q_{\textsf{odd}}$
and $Q_{\textsf{even}}$
\begin{equation}
Q_{\textsf{odd}}^2-Q_{\textsf{even}}^2=0\quad\text{and}\quad
[Q_{\textsf{odd}},\,Q_{\textsf{even}}]=0.
\end{equation}
Starting from this moment all the results obtained in Section~\ref{sec:newBRST}
can be applied to the BRST charge defined by \eqref{ber:55}
without any modifications. Moreover one can check that the ghost kinetic
operator \eqref{formal-q} satisfies the equation
$\{\hat{\eta}_0,\,\hat{\Qc}\}=0$.

\dopage{\finkfile}

\end{document}